\documentclass[aps,prl,twocolumn,superscriptaddress,showpacs,amsmath,amssymb]{revtex4-1}

\usepackage{amsfonts}
\usepackage{amssymb}
\usepackage{graphicx}
\usepackage{color}

\usepackage{bm}
\usepackage{braket}

\usepackage{hyperref}

\usepackage[normalem]{ulem}

\begin{document}
\title{Vanishing and non-vanishing persistent currents of various conserved quantities}
\author{Hirokazu Kobayashi}
\affiliation{Department of
Applied Physics, University of Tokyo, Tokyo 113-8656, Japan.}
\author{Haruki Watanabe} \email[]{hwatanabe@g.ecc.u-tokyo.ac.jp}
\affiliation{Department of
Applied Physics, University of Tokyo, Tokyo 113-8656, Japan.}

\begin{abstract}
For every conserved quantity written as a sum of local terms, there exists a corresponding current operator that satisfies the continuity equation.
The expectation values of current operators at equilibrium define the persistent currents that characterize spontaneous flows in the system.
In this work, we consider quantum many-body systems on a finite one-dimensional lattice and discuss the scaling of the persistent currents as a function of the system size.
We show that, when the conserved quantities are given as the Noether charges associated with internal symmetries or the Hamiltonian itself, the corresponding persistent currents can be bounded by a correlation function of two operators at a distance proportional to the system size, implying that they decay at least algebraically as the system size increases. 
In contrast, the persistent currents of accidentally conserved quantities can be nonzero even in the thermodynamic limit and even in the presence of the time-reversal symmetry. 
We discuss `the current of energy current' in $S=1/2$ XXZ spin chain as an example and obtain an analytic expression of the persistent current.
\end{abstract}

\maketitle

{\it Introduction.}---
The Noether theorem predicts the presence of a conserved quantity for every global continuous symmetry~\cite{Noether,Weinberg,Altland}. This fundamental theorem underlies the conservation of many important quantities such as the energy and the momentum in uniform stationary systems and the U(1) charges in many-body systems. There can also be other types of conserved quantities that commute with the Hamiltonian without apparent symmetry reasons. Such quantities are the key behind the integrability of exactly solvable models. They also affect the thermalization property of the system~\cite{Rigol1,Eckstein,Rigol2}.

\begin{figure}
\begin{center}
\includegraphics[width=\columnwidth]{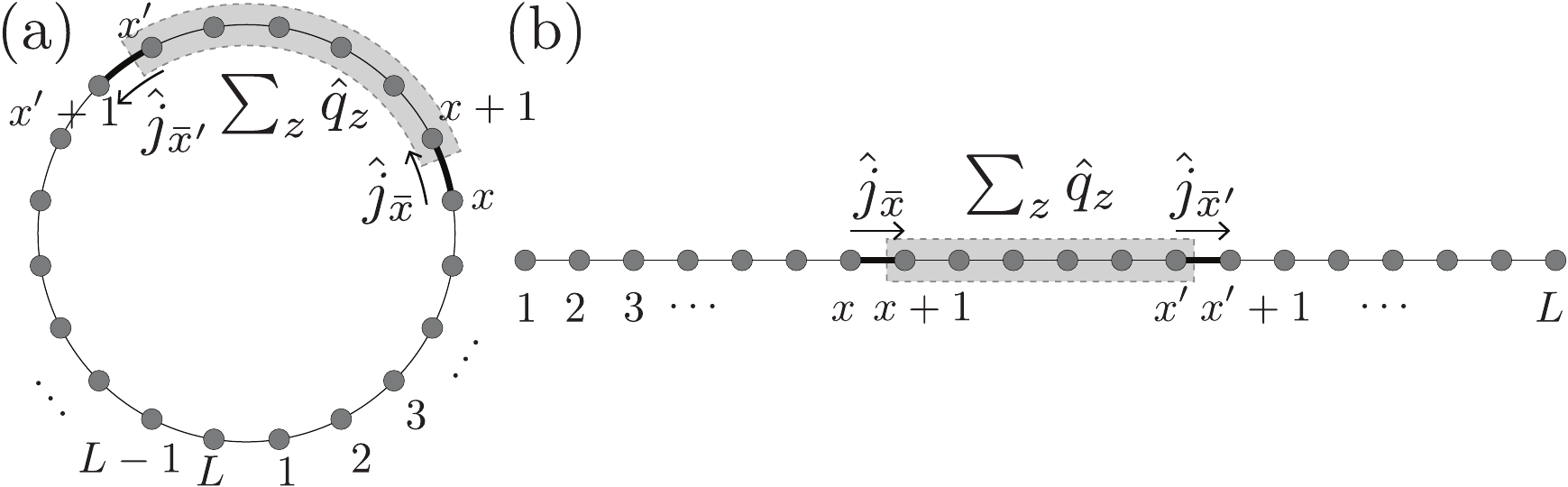}
\caption{\label{fig1} 
The lattice $\Lambda$ with (a) the periodic boundary condition and (b) the open boundary condition. 
The link between the sites $x$ and $x+1$ is denoted by $\bar{x}$.
}
\end{center}
\end{figure}

For each conserved quantity, one can define a current operator that satisfies the continuity equation [see Eq.~\eqref{continuity12} below]. We call the expectation value of current operators at equilibrium ``persistent currents."  Persistent currents can flow in systems which do not have any ends, such as the one-dimensional ring illustrated in Fig.~\ref{fig1}(a). 
Based on a variational argument that utilizes the so-called `twist operator,' Bloch showed that the persistent current for the U(1) charge vanishes in the limit of large system size in quasi one-dimensional systems~\cite{Bohm,Momoi,Yamamoto,Watanabe,Bachmann}. Recently, Kapustin and Spodyneiko proved a corresponding statement for the persistent energy current~\cite{Kapustin}. Their argument was rather a new approach focusing on the current response towards deformations of the Hamiltonian.

Then natural questions arise: do the persistent currents of other conserved quantities vanish in the thermodynamic limit, just like the persistent current of the U(1) charge and the energy? If so, how do we prove it? What are their scaling as a function of the system size? In this work, we answer these basic questions one by one. It is known that current operators have ambiguities in their definition. To address these questions in a meaningful manner, we should first show that the persistent current is independent of such ambiguity. 

Our analysis also provides an alternative proof of the absence of persistent energy current in the thermodynamic limit. As compared to the one in Ref.~\cite{Kapustin}, our argument is advantageous in two ways: (i) the finite-size scaling is accessible and (ii) the assumption of the absence of a finite-temperature phase transition~\cite{Kapustin} is not needed. Moreover, Bloch's original approach for the U(1) current only provides $O(L^{-1})$ bound regardless of the system size and the temperature~\cite{Bohm,Momoi,Yamamoto,Watanabe,Bachmann} but our argument improves it to an exponential decay at a finite temperature when the system size is large enough.

{\it Setting.}---
We consider a quantum many-body system defined on a finite one-dimensional lattice $\Lambda\equiv\{1,2,\dots,L\}$ ($L\in\mathbb{N}$). The boundary condition is set to be periodic so that the distance between two sites $x,y\in\Lambda$ is given by $d(x,y)\equiv \min_n|x-y+nL|$. For example, the site $x=L$ is right next to $x=1$ because $d(L,1)=1$ [see Fig.~\ref{fig1}(a)].  Hence, $x+nL$ ($n\in\mathbb{N}$) may be identified with $x\in\Lambda$.  The Hamiltonian of the system $\hat{H}\equiv\sum_{x=1}^L\hat{h}_x$ is given as the sum of local terms $\hat{h}_x$ that are Hermitian and are supported around $x$. The ranges of $\hat{h}_x$'s are bounded by a finite constant $r_h\in\mathbb{N}$.  %Namely, $\hat{h}_x$ does not affect local states at the site $y$ if $d(x,y)>r_h$. 
The system is not necessarily translation invariant. 

Suppose that there exists a Hermitian operator $\hat{Q}\equiv\sum_{x=1}^L\hat{q}_x$ that commutes with the Hamiltonian. The ranges of $\hat{q}_x$'s are also bounded by a finite constant $r_q\in\mathbb{N}$.  By definition, $[\hat{h}_x,\hat{q}_{y}]=0$ if $d(x,y)>r$ with $r\equiv r_h+r_q$. The system size $L$ is assumed to be much bigger than $r$.  When $\hat{q}_x$ generates a compact Lie group exclusively at site $x$ (i.e., the range $r_q=0$),  $\hat{q}_x$ must be integer-valued in some unit, which we set $1$ by a proper normalization. In such a case, the twist operator $\hat{U}\equiv\exp\big[2\pi iL^{-1}\sum_{x=1}^Lx\hat{q}_x\big]$ is well defined and Bloch's variational argument is applicable~\cite{Bohm,Momoi,Yamamoto,Watanabe,Bachmann}. Here we proceed without assuming such properties of $\hat{q}_x$.

Given $\hat{H}$ and $\hat{Q}$, we introduce a current operator associated with the link $\bar{x}$ between $x$ and $x+1$ that satisfies the continuity equation [see Fig.~\ref{fig1}(a)]
\begin{equation}
i[\hat{H},\sum_{z=x+1}^{x'}\hat{q}_z]=\hat{j}_{\bar{x}}-\hat{j}_{\bar{x}'}\quad (x'> x).\label{continuity12}
\end{equation}
We assume that $\hat{j}_{\bar{x}}$ is localized around the link $\bar{x}$ with a finite support. When $d(x,x')\gg r$, the supports of $\hat{j}_{\bar{x}}$ and $\hat{j}_{\bar{x}'}$ in the right hand side do not overlap and $\hat{j}_{\bar{x}}$ can be uniquely singled out, for the given $\hat{h}_x$ and $\hat{q}_x$. Note that $\langle\hat{j}_{\bar{x}}\rangle$ is independent of the link $\bar{x}$. This is because, for any operator $\hat{o}$, 
\begin{equation}
\langle[\hat{H},\hat{o}]\rangle=\langle\hat{H}\hat{o}\rangle-\langle\hat{o}\hat{H}\rangle=0, \label{HO}
\end{equation}
when the expectation value is computed with respect to the Gibbs state or the ground state (or any eigenstate) of $\hat{H}$.  We obtain $\langle\hat{j}_{\bar{x}}\rangle=\langle\hat{j}_{\bar{x}'}\rangle$ by applying Eq.~\eqref{HO} to Eq.~\eqref{continuity12}~\cite{Watanabe}. 

The decomposition of $\hat{H}$ and $\hat{Q}$ into local terms $\hat{h}_x$ and $\hat{q}_x$ is not unique~\cite{Kitaev}. Let
$\hat{H}=\sum_{x=1}^L\hat{h}_x'$ and $\hat{Q}=\sum_{x=1}^L\hat{q}_x'$ be an alternative decomposition, and let $\hat{j}_{\bar{x}}'$ be the current operator corresponding to this choice. Owing to the assumed locality of $\hat{q}_x$, we can write
\begin{equation}
\sum_{z=x+1}^{x'}\hat{q}_z'=\sum_{z=x+1}^{x'}\hat{q}_z+\delta\hat{q}_{\bar{x}}-\delta\hat{q}_{\bar{x}'}
\end{equation}
using operators $\delta\hat{q}_{\bar{x}}$ and $\delta\hat{q}_{\bar{x}'}$ localized around $\bar{x}$ and $\bar{x}'$, respectively. Substituting this into the continuity equation  Eq.~\eqref{continuity12}, we find $\hat{j}_{\bar{x}}'-\hat{j}_{\bar{x}}=i[\hat{H},\delta\hat{q}_{\bar{x}}]$. Again applying Eq.~\eqref{HO}, we conclude that $\langle\hat{j}_{\bar{x}}\rangle$ is independent of the choice of local operators, despite that the current operator itself may be ambiguous.  A similar conclusion can be found in Refs.~\cite{Doyon,Pozsgay1}, but our discussion here is slightly more general in that we assumed only the locality of $\hat{q}_x$.

\begin{figure}
\begin{center}
\includegraphics[width=\columnwidth]{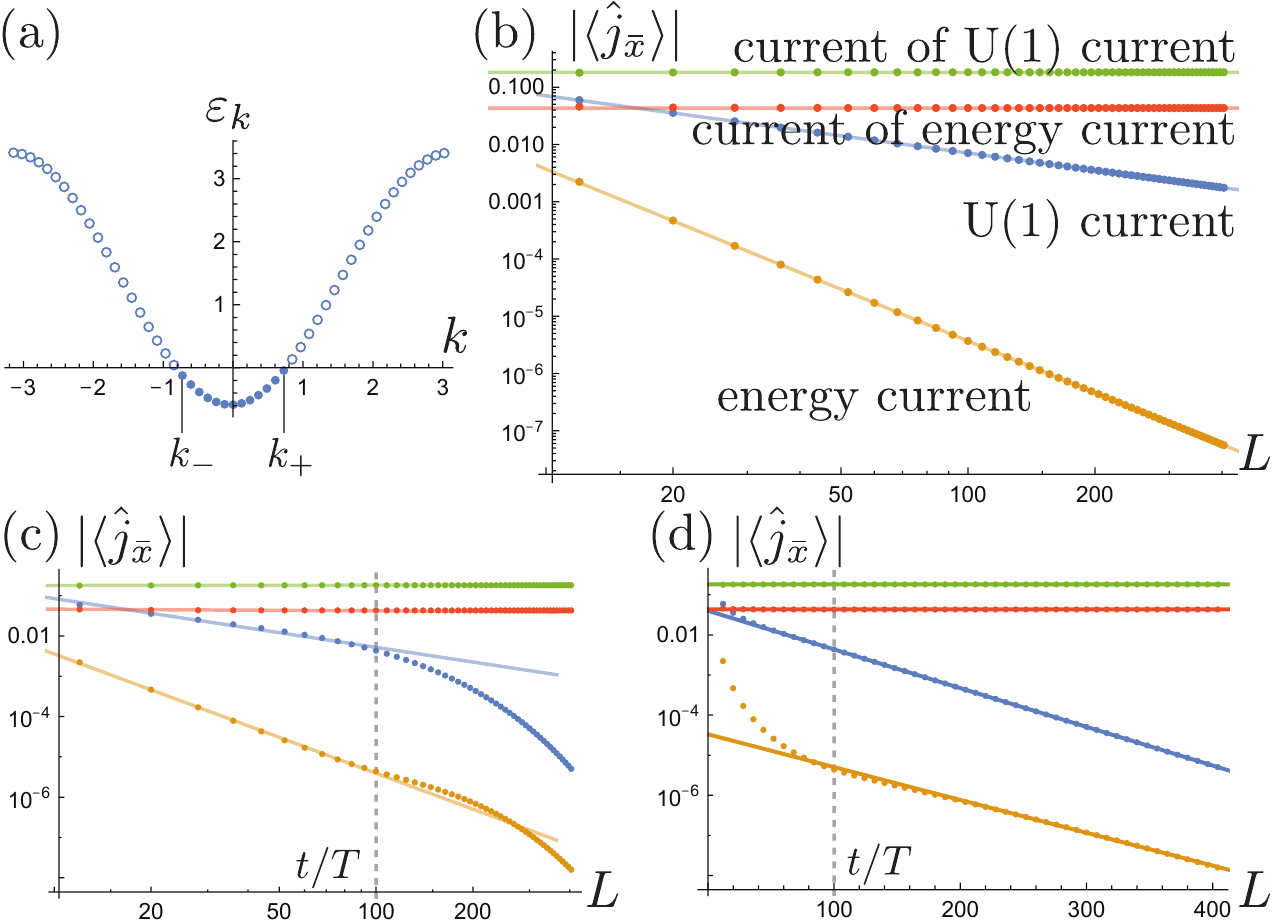}
\caption{\label{fig2} 
Persistent currents in the tight-binding model with $t_1=-te^{-i\theta/L}$ and $t_0=\mu/2$. We set $t=1$, $\mu=\sqrt{2}$ (corresponding to the quarter filling), $\theta=\pi/2$, and $L=8n-4$ ($n\in\mathbb{N}$).
(a) The band dispersion $\varepsilon_k$ for $L=52$.
(b) The log-log plot of the persistent currents at $T=0$.  Lines are obtained by fitting. 
The slopes for the U(1) current and the energy current corresponds to $L^{-1}$ and $L^{-3}$ decay.
(c,d) The log-log and linear-log plot of the persistent currents at $T=t/100$. We use the same color labels as in (b).
}
\end{center}
\end{figure}

{\it Tight-binding Model.}---
Before further presenting abstract arguments, let us first discuss illustrative examples. We first consider a single-band tight-binding model with hopping parameters $t_d\in\mathbb{C}$:
\begin{align}
\hat{H}=\sum_{x=1}^L\hat{h}_x,\quad\hat{h}_x=\sum_{d=0}^{r_h}t_d\hat{c}_{x+d}^\dagger \hat{c}_{x}+\text{h.c.},
\end{align}
where $\hat{c}_{x}$ is the annihilation operator of fermions at site $x\in\Lambda$.   Introducing the Fourier transformation $\hat{c}_k^\dagger \equiv L^{-1/2}\sum_{x=1}^L\hat{c}_x^\dagger e^{ikx}$ for $k=2\pi j/L$, we obtain the diagonalized form $\hat{H}=\sum_k\varepsilon_k\hat{c}_{k}^\dagger\hat{c}_{k}$. The band dispersion $\varepsilon_k\equiv\sum_{d=0}^{r_h}t_d e^{-ikd}+\text{c.c.}$ defines the group velocity $v_k\equiv\partial_k\varepsilon_k=-i\sum_{d=0}^{r_h}dt_d e^{-ikd}+\text{c.c}$. The ground state in a $N$ fermion system is given by the Slater determinant of the $N$-lowest energy states. We fix the filling $\nu=N/L$ in the canonical ensemble, while $N$ will be automatically chosen by fully occupying states with $\varepsilon_k<0$ in the ground canonical ensemble (the chemical potential $\mu$ is included in $\varepsilon_k$ via $t_0=\mu/2$).
For brevity, here we assume that the ground state is unique and that states with momentum $k$ in the range $k_-\leq k\leq k_+$ are occupied and those outside are unoccupied in the ground state [see Fig.~\ref{fig2}(a)].

Let us consider a Hermitian operator of the form 
\begin{align}
\hat{Q}=\sum_{x=1}^L\hat{q}_x,\quad \hat{q}_x=\sum_{d'=0}^{r_q}q_{d'}\hat{c}_{x+d'}^\dagger \hat{c}_{x}+\text{h.c.}, \label{qTB}
\end{align}
which commutes with the Hamiltonian as it is diagonal in the Fourier space: $\hat{Q}=\sum_{k}q_k\hat{c}_{k}^\dagger \hat{c}_{k}$ with $q_k\equiv \sum_{d'=0}^{r_q}q_{d'} e^{-ikd'}+\text{c.c}$. For example, $q_k=1$ for the U(1) charge $\hat{q}_x=\hat{c}_x^\dagger\hat{c}_x$ and $q_k=\varepsilon_k$ for the energy $\hat{q}_x=\hat{h}_x$.
Using the continuity equation \eqref{continuity12}, we identify the current operator $\hat{j}_{\bar{x}}=\sum_{d=1}^{r_h}\sum_{d''=1}^{d}\hat{\sigma}_{x-d''+1}^{d}$, where
\begin{align}
\hat{\sigma}_{x}^{d}\equiv\sum_{d'=0}^{r_q}i(t_{d}^*q_{d'}^*\hat{c}_{x}^\dagger  \hat{c}_{x+d+d'}+t_{d}^*q_{d'}\hat{c}_{x+d'}^\dagger \hat{c}_{x+d})+\text{h.c.}
\end{align}
See Supplemental Material (SM) for the derivation. The averaged current operator $\hat{\bar{j}}\equiv L^{-1}\sum_{x=1}^L\hat{j}_{\bar{x}}$ takes an intuitive form $\hat{\bar{j}}=L^{-1}\sum_{k}q_kv_k\hat{c}_{k}^\dagger \hat{c}_{k}$ in the Fourier space, which is simply the charge $q_k$ multiplied by the group velocity $v_k=\partial_k\varepsilon_k$.  
The persistent current $\langle\hat{j}_{\bar{x}}\rangle=L^{-1}\sum_{k=k_-}^{k_+}q_kv_k$ at $T=0$ can thus be evaluated by the Euler--Maclaurin formula:
\begin{align}
\langle\hat{j}_{\bar{x}}\rangle=\int_{k_-}^{k_+}\frac{q_kv_k}{2\pi}dk+\frac{q_{k_+}v_{k_+}+q_{k_-}v_{k_-}}{2L}+O(L^{-2}).
\label{EMf}
\end{align}

Now we show that $\langle\hat{j}_{\bar{x}}\rangle$ vanishes in the large $L$ limit when $q_k$ is a function of $\varepsilon_k$ but not a function of $v_k$. This is the case when $\hat{Q}$ is the U(1) charge and the Hamiltonian itself. If we write $q_k=f'(\varepsilon_k)$ [$f(\varepsilon_k)=\varepsilon_k$ for the U(1) charge and $f(\varepsilon_k)=\varepsilon_k^2/2$ for the energy], the first term in Eq.~\eqref{EMf} can be written as $[f(\varepsilon_{k_+})-f(\varepsilon_{k_-})]/(2\pi)$, which is small because $|\varepsilon_{k_+}-\varepsilon_{k_-}|=O(L^{-1})$ in the ground state. 
In the canonical ensemble, $\varepsilon_{k_\pm}=O(1)$ and $\langle\hat{j}_{\bar{x}}\rangle\propto L^{-1}$ in general.
In the ground canonical ensemble, $\varepsilon_{k_\pm}$ themselves are $O(L^{-1})$ and the persistent energy current decays faster.

On the other hand, in more general cases, $q_k$ may not take the above form. 
For example, we can re-use the above current operator $\hat{j}_{\bar{x}}$ as an example of $\hat{q}_x$ in Eq.~\eqref{qTB}. Then the corresponding $q_k$ is $v_k$ for the U(1) current and $\varepsilon_kv_k$ for the energy current.  In this case, the first term in Eq.~\eqref{EMf} does not vanish in general.
We demonstrate these results in Fig.~\ref{fig2}(b) using the simplest model with the nearest neighbor hopping.
We also show the result for a finite temperature $T=t/100$ in Fig.~\ref{fig2}(c,d), which demonstrates the crossover to the exponential decay around $L\sim t/T$.

{\it XXZ Spin Chain.}---The above discussion heavily relies on the simplicity of the noninteracting model. However, the key conclusion remains valid even in the presence of  interactions. As an example, let us consider the $S=1/2$ XXZ spin chain. The Hamiltonian is  $\hat{H}=\sum_{x=1}^{L}\hat{h}_x$ with
\begin{align}
\hat{h}_x=J^2\left(\frac{1}{2}\hat{s}_{x+1}^{+}\hat{s}_{x}^{-}+\frac{1}{2}\hat{s}_{x+1}^{-}\hat{s}_{x}^{+}+\Delta \hat{s}_{x+1}^{z}\hat{s}_{x}^{z}\right),
\end{align}
where $\hat{s}_{x}^{\pm}=\hat{s}_{x}^x\pm i\hat{s}_{x}^y$ and $\hat{s}_x^z$ are the spin $1/2$ operators at site $x\in\Lambda$.  
The energy current operator  for $\hat{q}_x=\hat{h}_x$ is given by $\hat{j}_{\bar{x}}=i[\hat{h}_x,\hat{h}_{x+1}]$~\cite{PhysRevB.55.11029}.
The total energy current $\hat{Q}=\sum_{x=1}^L\hat{j}_{\bar{x}}$ commutes with the Hamiltonian~\cite{PhysRevB.55.11029}, allowing us to discuss `the current of the energy current'~\cite{Joel1,Joel2,Pozsgay3}. We append the concrete expressions of these operators in SM. Since the energy current are odd under the time-reversal symmetry, `the current of energy current' is even. Thus, it can flow even in the presence of the time-reversal symmetry.  In contrast, the total U(1) current $\hat{Q}=\sum_{x=1}^Li[\hat{h}_x,\hat{s}_{x+1}^z]$ corresponding to $\hat{q}_x=\hat{s}_x^z$ and `the total current of energy current'  do not commute with Hamiltonian unless $\Delta=0$. 

\begin{figure}
\begin{center}
\includegraphics[width=\columnwidth]{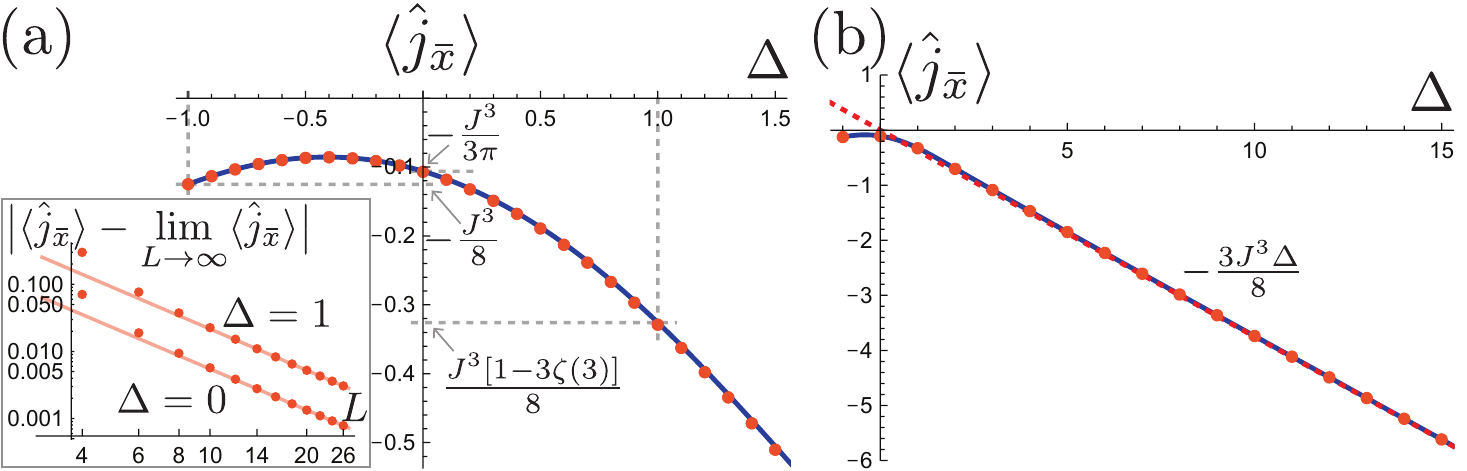}
\caption{\label{fig3}
The expectation value of `the current of the energy current' in the XXZ spin chain with $J=1$. Panels (a) and (b) display different ranges of $\Delta$.
Orange points are obtained by exact diagonalization for the $L=26$ chain. 
The blue solid curves are the analytic expression in Eq.~\eqref{Jxxz}  in the thermodynamic limit.
The red dashed line in (b) represents the asymptotic behavior as $\Delta\rightarrow+\infty$ in the thermodynamic limit.
The inset in (a) checks the convergence as $L$ increases for $\Delta=0$ and $1$. The fitting lines have the slope corresponding to the $L^{-2}$ correction.
}
\end{center}
\end{figure}

The $\Delta=0$ point reduces the tight-binding model with $t_1=-J/2$ and $k_{\pm}=\pm\pi/2$ and Eq.~\eqref{EMf} gives $\langle\hat{j}_{\bar{x}}\rangle=-J^3/(3\pi)$ in the large $L$ limit.  
For more general values of $\Delta$, $\langle\hat{j}_{\bar{x}}\rangle$ can be expressed in terms of correlation functions of up to four neighboring spin operators. Using the results of Ref.~\cite{Kato}, we find the following expressions for $-1<\Delta\leq1$ in the large $L$ limit (see SM):
\begin{align}
&\langle\hat{j}_{\bar{x}}\rangle=J^3\frac{\pi \Delta-2(1-\Delta^2)^{3/2}\zeta_{\eta}(3)}{8\pi},\label{Jxxz}\\
&\zeta_{\eta}(3)\equiv\int_{-\infty}^{+\infty}\frac{1}{\sinh (x-i\frac{\pi}{2})}\frac{\cosh[\eta (x-i\frac{\pi}{2})]}{\sinh^3[\eta (x-i\frac{\pi}{2})]}dx,\label{Jxxz2}
\end{align}
where $\eta$ ($0\leq\eta<1$) is defined by $\Delta=\cos(\pi\eta)$.  For example, we find $\langle\hat{j}_{\bar{x}}\rangle=J^3[1-3\zeta(3)]/8$ at $\Delta=1$ [$\zeta(z)$ is the Riemann zeta function] and $\langle\hat{j}_{\bar{x}}\rangle\to-J^3/8$ as $\Delta\to-1$.
The expressions in Eqs.~\eqref{Jxxz} and \eqref{Jxxz2} are also valid for $\Delta>1$, for which we set $\eta=i\eta'$ with $\eta'>0$. For example, we have $\langle\hat{j}_{\bar{x}}\rangle\simeq-3J^3\Delta/8$ when $\Delta\gg1$.
Our numerical results up to $26$ spins are presented in Fig.~\ref{fig3}.

{\it Vanishing Persistent Current under Open Boundary Condition.}---We have seen through examples that not all persistent currents vanish in the large $L$ limit: the current associated with internal symmetries and the energy are $O(L^{-1})$ but the current for accidentally conserved quantities are $O(L^{0})$. To have a better understanding on these differences, here we temporarily consider the open boundary condition (OBC) and prove that the persistent current vanishes for any conserved quantity under OBC. This consideration serves as the reference point when evaluating the persistent current under periodic boundary condition (PBC) later.

The crucial difference between PBC and OBC lies in the definition of the distance.  Under OBC, the distance between two sites $x,y\in\Lambda$ is simply given by $\tilde{d}(x,y)\equiv |x-y|$. Thus the sites $x=L$ and $x=1$ are at a long distance [see Fig.~\ref{fig1}(b)]. Let $\hat{\tilde{H}}\equiv\sum_{x=1}^L\hat{\tilde{h}}_x$ be the Hamiltonian under OBC, in which all interactions across the `seam' between $x=L$ and $x=1$ are switched off.
We demand that the local Hamiltonians remain unchanged in the bulk, i.e., $\hat{\tilde{h}}_x=\hat{h}_x$ when $r_h< x<L-r_h+1$. Near boundaries (i.e., $1\leq x\leq r_h$ or $L-r_h+1\leq x\leq L$), $\hat{\tilde{h}}_x$'s are arbitrary as long as $\hat{\tilde{h}}_x$ is supported around $x$ and its range is bounded by $r_h$.

Another important assumption on $\hat{\tilde{H}}$ is that there exists a conserved charge $\hat{\tilde{Q}}\equiv\sum_{x=1}^L\hat{\tilde{q}}_x$ that may differ from $\hat{Q}$ only near the boundaries. 
For example, in the case of internal symmetries, one can set $\hat{\tilde{Q}}=\hat{Q}$ by symmetrizing $\hat{\tilde{H}}$ using $e^{i\theta\hat{Q}}$. In contrast, accidentally conserved quantities of $\hat{H}$ may not have a correspondence in $\hat{\tilde{H}}$. For example, the total energy current in the XXZ model is not conserved under OBC~\cite{Grabowski_1996}.

We assume that the current operator $\hat{\tilde{j}}_{\bar{x}}$, satisfying the continuity equation
\begin{align}
&i[\hat{\tilde{H}},\sum_{z=x+1}^{x'}\hat{\tilde{q}}_z]=\hat{\tilde{j}}_{\bar{x}}-\hat{\tilde{j}}_{\bar{x}'}\quad (x'> x),\label{continuity22}
\end{align}
remains localized around the link $\bar{x}$ with a finite support.  If we set $x=0$ and $x'=L$ in Eq.~\eqref{continuity22}, we find $\hat{\tilde{j}}_{\bar{L}}-\hat{\tilde{j}}_{\bar{0}}=-i[\hat{\tilde{H}},\hat{\tilde{Q}}]=0$.
Because of the assumed locality of $\hat{\tilde{j}}_{\bar{0}}$ and $\hat{\tilde{j}}_{\bar{L}}$, this is equivalent with $\hat{\tilde{j}}_{\bar{0}}=\hat{\tilde{j}}_{\bar{L}}=0.$ This is reasonable since nothing can flow into or flow out of the system under OBC. Then, again from Eq.~\eqref{continuity22}, we find a compact expression of the current operator $\hat{\tilde{j}}_{\bar{x}}=\sum_{z=x+1}^{L}i[\hat{\tilde{H}},\hat{\tilde{q}}_z]$,  which takes the form of $[\hat{\tilde{H}},\hat{o}]$.
Thus the persistent current, computed using the Gibbs state or the ground state of $\hat{\tilde{H}}$, is precisely zero under OBC without the large $L$ limit. 

{\it Interpolating Hamiltonian.}---
Now, we return to the PBC in which the distance is measured by $d(x,y)$. The Hamiltonian $\hat{\tilde{H}}$ for OBC can be regarded as the Hamiltonian under PBC, since it satisfies the locality condition. In contrast, the Hamiltonian $\hat{H}$ for PBC cannot be used under OBC in general, since it may contain interactions between the two boundaries $x=L$ and $1$ that are regarded as long-ranged with respect to $\tilde{d}(x,y)$ of OBC.

We introduce a one-parameter family of Hamiltonians $\hat{H}(s)=\sum_{x=1}^L\hat{h}_x(s)\equiv s\hat{H}+(1-s)\hat{\tilde{H}}$ for $s\in[0,1]$, which linearly interpolates our original Hamiltonian $\hat{H}(1)=\hat{H}$ and the reference Hamiltonian $\hat{H}(0)=\hat{\tilde{H}}$. By construction of $\hat{\tilde{H}}$, the local Hamiltonians $\hat{h}_x(s)$ in the `bulk' region does not depend on $s$, i.e., 
\begin{equation}
\hat{h}_x(s)=\hat{h}_x,\quad r_h< x<L-r_h+1. \label{hOBCPBC}
\end{equation}
In the following, we denote by $\langle \hat{o}\rangle_s$ the expectation value with respect to the Gibbs state $\hat{\rho}(s)\equiv Z(s)^{-1}e^{-\hat{H}(s)/T}$ [$Z(s)\equiv\text{tr}\,e^{-\hat{H}(s)/T}$] at a finite $T$ or the ground state of $\hat{H}(s)$ at $T=0$. 

We assume that, for any $s\in[0,1]$, the system has a conserved charge $\hat{Q}(s)\equiv\sum_{x=1}^L\hat{q}_x(s)$ with $\hat{Q}(1)=\hat{Q}$ and $\hat{Q}(0)=\hat{\tilde{Q}}$, 
in which $\hat{q}_x(s)$ is independent of $s$ at least when $x$ is away from $1$ and $L$:
\begin{equation}
\hat{q}_x(s)=\hat{q}_x\quad \text{if}\quad 1\ll x \ll L.\label{rOBCPBC}
\end{equation}
These assumptions are automatically fulfilled for the case of the Hamiltonian $\hat{Q}(s)=\hat{H}(s)$. 
Also, for internal symmetries with $\hat{Q}=\hat{\tilde{Q}}$, we can set $\hat{q}_x(s)=\hat{q}_x$ for any $x\in\Lambda$ and $s\in[0,1]$.  
Substituting Eqs.~\eqref{hOBCPBC} and \eqref{rOBCPBC} into the continuity equation
\begin{align}
i[\hat{H}(s),\sum_{z=x+1}^{x'}\hat{q}_z(s)]+\hat{j}_{\bar{x}'}(s)-\hat{j}_{\bar{x}}(s)=0\,\,\,\,(x'> x),\label{continuity32}
\end{align}
we find that the current operator $\hat{j}_{\bar{x}}(s)$ is also independent of $s$ when $x$ is away from $1$ and $L$.

\begin{figure}
\begin{center}
\includegraphics[width=\columnwidth]{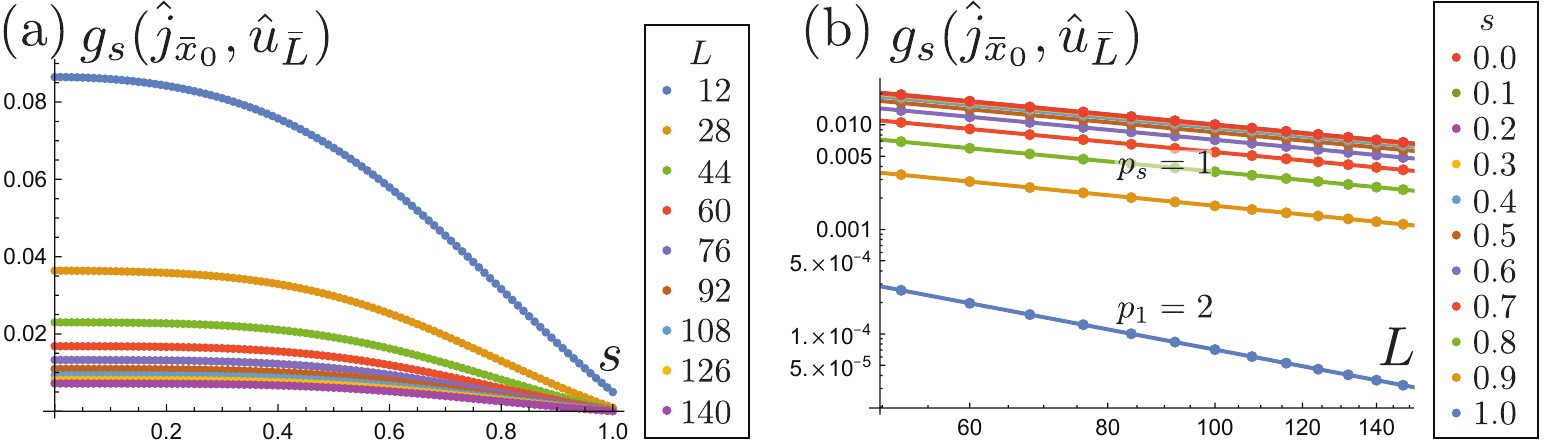}
\caption{\label{fig4} 
$g_s(\hat{j}_{\bar{x}_0},\hat{u}_{\bar{L}})$ for the U(1) current in the tight-binding model in Fig.~\ref{fig2}. We set $t=1$ and $\theta=\pi/2$. The lines in the panel (b) are obtained by fitting.
The power $p_s$ is $1$ for $0\leq s<1$ except $p_1=2$. 
}
\end{center}
\end{figure}

{\it Bound for the Persistent Current}.---
With these preparations, let us evaluate $\langle \hat{j}_{\bar{x}}\rangle$ for the original Hamiltonian $\hat{H}$.
Let $x_0\in\Lambda$ be a site away from $x=1$ and $L$. We have
\begin{align}
\langle \hat{j}_{\bar{x}}\rangle&=\langle \hat{j}_{\bar{x}_0}\rangle_{s=1}=\int_{0}^1 ds\,\partial_s\langle \hat{j}_{\bar{x}_0}\rangle_s,
\label{integral}
\end{align}
where we used the fact that $\langle\hat{j}_{\bar{x}}\rangle$ is independent of $\bar{x}$ and that $\hat{j}_{\bar{x}_0}(s)$ is $s$ independent as long as $1\ll x_0\ll L$. The last equality is the absence of the persistent current under OBC as discussed above. 

According to the linear response theory, the derivative $\partial_s\langle \hat{j}_{\bar{x}_0}\rangle_{s}$ is given by a correlation function $\partial_s\langle \hat{j}_{\bar{x}_0}\rangle_{s}=-g_s(\hat{j}_{\bar{x}_0},\hat{u}_{\bar{L}})$, 
where
$\hat{u}_{\bar{L}}\equiv\partial_s\hat{H}(s)=\sum_{x=1}^L(\hat{h}_x-\hat{\tilde{h}}_x)$
is localized around the link $\bar{L}$ between $x=L$ and $x=1$.  
At a finite $T$, $g_s(\hat{o},\hat{o}')=T^{-1}\langle\langle\hat{o},\hat{o}'\rangle\rangle_s$ is given by is the canonical correlation
\begin{align}
\langle\langle\hat{o},\hat{o}'\rangle\rangle_s&\equiv T\int_{0}^{T^{-1}}d\alpha\big\langle e^{\alpha\hat{H}(s)}\hat{o}e^{-\alpha\hat{H}(s)}\hat{o}'\big\rangle_s-\langle \hat{o}\rangle_s\langle\hat{o}'\rangle_s,
\end{align}
and, at $T=0$,
\begin{equation}
g_s(\hat{o},\hat{o}')=\Big\langle\hat{o}\frac{1-\hat{P}(s)}{\hat{H}(s)-E_0(s)}\hat{o}'\Big\rangle_s+\text{c.c.},
\end{equation}
where $\hat{P}(s)$ is the projection onto the ground state of $\hat{H}(s)$ and $E_0(s)$ is the ground state energy.  Because of the property $g_s([\hat{H}(s),\hat{o}],\hat{o}')=\langle[\hat{o}',\hat{o}]\rangle_s$, $g_s(\hat{j}_{\bar{x}_0},\hat{u}_{\bar{L}})$ is independent of $x_0$ as long as $1\ll x_0\ll L$. Up to this point, all expressions are exact.

Now, recall that $\hat{j}_{\bar{x}_0}$ and $\hat{u}_{\bar{L}}$ are respectively localized around the link $\bar{x}_0$ and $\bar{L}$. If  $x_0$ is set to be $L/2$ when $L$ is even and $(L+1)/2$ when $L$ is odd, the distance between the supports of these two operators can be approximated by $L/2$.  Hence, even in gapless systems, $g_s(\hat{j}_{\bar{x}_0},\hat{u}_{\bar{L}})$ should decay as $L$ increases. 
Let us postulate the power-law decay, i.e., $|g_s(\hat{j}_{\bar{x}_0},\hat{u}_{\bar{L}})|<c_sL^{-p_s}$ first. For example, Fig.~\ref{fig3} illustrates the case for the above tight-biding model at $T=0$. In this case, the persistent current can be bounded as
\begin{align}
|\langle \hat{j}_{\bar{x}}\rangle|\leq \int_0^1ds|g_s(\hat{j}_{\bar{x}_0},\hat{u}_{\bar{L}})|\leq cL^{-p}
\label{bound1}
\end{align}
with $c\equiv\max_sc_s$ and $p\equiv\min_s p_s$.
In contrast, when the correlation function decays exponentially, i.e., $|g_s(\hat{j}_{\bar{x}_0},\hat{u}_{\bar{L}})|\leq c_s'e^{-L/\xi_s}$, we instead have
\begin{align}
|\langle \hat{j}_{\bar{x}}\rangle|\leq \int_0^1ds|g_s(\hat{j}_{\bar{x}_0},\hat{u}_{\bar{L}})|\leq  c'e^{-L/\xi}
\label{bound2}
\end{align}
with $c'\equiv\max_sc_s'$ and $\xi\equiv\max_s \xi_s$. 
In gapless systems at a finite temperature, if the gapless mode has the velocity $v$, a crossover from the algebraic-decay regime ($L\lesssim v /T$) to the exponential-decay regime ($L\gtrsim v/T$) is expected in general~\cite{giamarchi,korepin}, as we have seen in Fig.~\ref{fig2}(c,d).  

{\it Conclusions.}---
In this work, we considered a process in which all interactions across the seam between $x=L$ and $x=1$ are gradually switched off. 
When the quantity $\hat{Q}$, possibly modified in accordance with the Hamiltonian, remains conserved during this process, the persistent current can be bounded by a correlation function of two operators separated by $L/2$ as in Eqs.~\eqref{bound1} and \eqref{bound2}. This is the case for internal symmetries and the Hamiltonian itself. In contrast, when $\hat{Q}$ fails to be conserved during the process, this argument is not applicable and the persistent current can be nonzero even in the large $L$ limit. The mechanism for nonvanishing persistent currents here is different from the anomalous mechanism proposed recently~\cite{Dominic,Watanabe2}.
Although our discussion was limited to (quasi) one-dimensional systems, several implications on higher dimensional systems can be derived in the same way as in Ref.~\cite{Watanabe}.

\begin{acknowledgements}
We would like to thank Yohei Fuji, Hosho Katsura, Masaki Oshikawa, and Hal Tasaki for useful discussions.
The work of H.W. is supported by JSPS KAKENHI Grant No.~JP20H01825 and by JST PRESTO Grant No.~JPMJPR18LA. 
\end{acknowledgements}

\bibliography{ref}

%\end{document}

\clearpage

\onecolumngrid
\appendix
\section{A. Current operators in the tight-binding model}
\label{app1}
The Hamiltonian $\hat{H}$ and the conserved charge $\hat{Q}$ are given by
\begin{align}
\hat{H}=\sum_{x=1}^L\hat{h}_x=\sum_{k}\varepsilon_k\hat{c}_{k}^\dagger \hat{c}_{k},\quad \hat{h}_x=\sum_{d=0}^{r_h}(t_d^*\hat{c}_{x}^\dagger \hat{c}_{x+d}+t_d\hat{c}_{x+d}^\dagger \hat{c}_{x}),\quad \varepsilon_k=\sum_{d=0}^{r_h}(t_d^* e^{ikd}+t_d e^{-ikd}),\\
\hat{Q}=\sum_{x=1}^L\hat{q}_x=\sum_{k}q_k\hat{c}_{k}^\dagger \hat{c}_{k},\quad \hat{q}_x=\sum_{d'=0}^{r_q}(q_{d'}^*\hat{c}_{x}^\dagger \hat{c}_{x+d'}+q_{d'}\hat{c}_{x+d'}^\dagger \hat{c}_{x}),\quad q_k=\sum_{d'=0}^{r_q}(q_{d'}^* e^{ikd'}+q_{d'} e^{-ikd'}).
\end{align}
The continuity equation reads
\begin{align}
\hat{j}_{\bar{x}-1}-\hat{j}_{\bar{x}}&=i[\hat{H},\hat{q}_x]=i\sum_{y=1}^L\sum_{d=0}^{r_h}\sum_{d'=0}^{r_q}[t_d^*\hat{c}_{y}^\dagger \hat{c}_{y+d}+t_d\hat{c}_{y+d}^\dagger \hat{c}_{y},q_{d'}^*\hat{c}_{x}^\dagger \hat{c}_{x+d'}+q_{d'}\hat{c}_{x+d'}^\dagger \hat{c}_{x}]\notag\\
&=\sum_{d=0}^{r_h}\sum_{d'=0}^{r_q}\left[i(t_{d}^*q_{d'}^*\hat{c}_{x-d}^\dagger \hat{c}_{x+d'}-t_{d}q_{d'}\hat{c}_{x+d'}^\dagger \hat{c}_{x-d})+i(t_{d}^*q_{d'}\hat{c}_{x+d'-d}^\dagger \hat{c}_{x}-t_{d}q_{d'}^*\hat{c}_{x}^\dagger \hat{c}_{x+d'-d})\right]\notag\\
&\quad-\sum_{d=0}^{r_h}\sum_{d'=0}^{r_q}\left[i(t_{d}^*q_{d'}^*\hat{c}_{x}^\dagger  \hat{c}_{x+d+d'}-t_{d}q_{d'}\hat{c}_{x+d+d'}^\dagger \hat{c}_{x})+i(t_{d}^*q_{d'}\hat{c}_{x+d'}^\dagger \hat{c}_{x+d}-t_{d}q_{d'}^*\hat{c}_{x+d}^\dagger  \hat{c}_{x+d'})\right]\notag\\
&\equiv\sum_{d=0}^{r_h}(\hat{\sigma}_{x-d}^{d}-\hat{\sigma}_{x}^{d}).
\end{align}
where $\hat{\sigma}_{x}^{d}$ is given in Eq.~(6) of the main text. Therefore, the current operator can be identified as
\begin{align}
&\hat{j}_{\bar{x}}=\sum_{d=1}^{r_h}\sum_{d''=1}^{d}\hat{\sigma}_{x-d''+1}^{d},\\
&\hat{\bar{j}}=\frac{1}{L}\sum_{d=0}^{r_h}d\sum_{x=1}^L\hat{\sigma}_{x}^{d}=\frac{1}{L}\sum_{k}\sum_{d=0}^{r_h}id(t_{d}^*e^{ikd}-t_{d}e^{-ikd})\sum_{d'=0}^{r_q}(q_{d'}^*e^{ikd'}+q_{d'}e^{-ikd'})\hat{c}_{k}^\dagger \hat{c}_{k}=\frac{1}{L}\sum_{k}q_k\partial_k\varepsilon_k\hat{c}_{k}^\dagger \hat{c}_{k}.
\end{align}

\section{B. The current of energy current in the XXZ spin chain}
\label{app2}
The Hamiltonian $\hat{H}$ and the energy current operator $\hat{j}_{\bar{x}}^{\text{E}}$ are given by 
\begin{align}
\hat{H}&=\sum_{x=1}^{L}\hat{h}_x,\quad\hat{h}_x=J\left(\frac{1}{2}\hat{s}_{x+1}^{+}\hat{s}_{x}^{-}+\frac{1}{2}\hat{s}_{x+1}^{-}\hat{s}_{x}^{+}+\Delta \hat{s}_{x+1}^{z}\hat{s}_{x}^{z}\right),\\
\hat{j}_{\bar{x}}^{\text{E}}&=i[\hat{h}_x,\hat{h}_{x+1}]=\frac{iJ^2}{2}\big(\hat{s}_{x+2}^{+}\hat{s}_{x+1}^z\hat{s}_{x}^{-}-\Delta\hat{s}_{x+2}^{z}\hat{s}_{x+1}^{+}\hat{s}_{x}^{-}-\Delta\hat{s}_{x+2}^{+}\hat{s}_{x+1}^{-}\hat{s}_{x}^{z}\big)+\text{h.c.}
\end{align}
The current operator of the energy current, i,e, the operator $\hat{j}_{\bar{x}}$ satisfying the continuity equation (1) for $\hat{q}_x=\hat{j}_{\bar{x}}^{\text{E}}$, is
\begin{align}
\hat{j}_{\bar{x}}&=\frac{J^3}{8}\left(\hat{s}^{+}_{x+2}\hat{s}^{-}_{x+1}-4\hat{s}^{+}_{x+3}\hat{s}^{z}_{x+2}\hat{s}^{z}_{x+1}\hat{s}^{-}_{x}\right)+\text{h.c.}+\frac{J^3\Delta^2}{8}\left(\hat{s}^{+}_{x+2}\hat{s}^{-}_{x+1}-4\hat{s}^{z}_{x+3}\hat{s}^{+}_{x+2}\hat{s}^{-}_{x+1}\hat{s}^{z}_{x}\right)+\text{h.c.}\notag\\
  &\quad+\frac{J^3\Delta}{4}\left(2\hat{s}^{+}_{x+3}\hat{s}^{z}_{x+2}\hat{s}^{-}_{x+1}\hat{s}^{z}_{x}+2 \hat{s}^{z}_{x+3}\hat{s}^{+}_{x+2}\hat{s}^{z}_{x+1}\hat{s}^{-}_{x}-\hat{s}^{+}_{x+3}\hat{s}^{-}_{x+2}\hat{s}^{+}_{x+1}\hat{s}^{-}_{x}+\hat{s}^{+}_{x+3}\hat{s}^{-}_{x+2}\hat{s}^{-}_{x+1}\hat{s}^{+}_{x}+\hat{s}^{z}_{x+2}\hat{s}^{z}_{x+1}\right)+\text{h.c.}
\end{align}
Generalizations to the XYZ spin chain and to the XXZ model with U(1) flux are straightforward. A similar expression was presented in Sec.~2.1 of Ref.~\cite{Pozsgay3} but there were several typos in coefficients and superscripts.

The expectation value is given in terms of the equal-time correlation functions:
\begin{align}
&\langle\hat{j}_{\bar{x}}\rangle/J^3\notag\\
&=\frac{1}{8}\left[\langle\hat{s}^{+}_{2}\hat{s}^{-}_{1}+\hat{s}^{+}_{1}\hat{s}^{-}_{2}\rangle-4(\langle\hat{s}^{+}_{4}\hat{s}^{z}_{3}\hat{s}^{z}_{2}\hat{s}^{-}_{1}\rangle+\langle\hat{s}^{+}_{1}\hat{s}^{z}_{2}\hat{s}^{z}_{3}\hat{s}^{-}_{4}\rangle)\right]\notag\\
&\quad+\frac{\Delta^2}{8}\left[\langle\hat{s}^{+}_{2}\hat{s}^{-}_{1}+\hat{s}^{+}_{1}\hat{s}^{-}_{2}\rangle-4(\langle\hat{s}^{z}_{4}\hat{s}^{+}_{3}\hat{s}^{-}_{2}\hat{s}^{z}_{1}\rangle+\langle\hat{s}^{z}_{1}\hat{s}^{+}_{2}\hat{s}^{-}_{3}\hat{s}^{z}_{4}\rangle)\right]\notag\\
&\quad+\frac{\Delta}{2}\left[(\langle\hat{s}^{+}_{4}\hat{s}^{z}_{3}\hat{s}^{-}_{2}\hat{s}^{z}_{1}\rangle+\langle\hat{s}^{z}_{1}\hat{s}^{+}_{2}\hat{s}^{z}_{3}\hat{s}^{-}_{4}\rangle)+(\langle\hat{s}^{z}_{4}\hat{s}^{+}_{3}\hat{s}^{z}_{2}\hat{s}^{-}_{1}\rangle+\langle\hat{s}^{+}_{1}\hat{s}^{z}_{2}\hat{s}^{-}_{3}\hat{s}^{z}_{4}\rangle)\right]\notag\\
&\quad+\frac{\Delta}{4}\left[-(\langle\hat{s}^{+}_{4}\hat{s}^{-}_{3}\hat{s}^{+}_{2}\hat{s}^{-}_{1}\rangle+\langle\hat{s}^{+}_{1}\hat{s}^{-}_{2}\hat{s}^{+}_{3}\hat{s}^{-}_{4}\rangle)+(\langle\hat{s}^{+}_{4}\hat{s}^{-}_{3}\hat{s}^{-}_{2}\hat{s}^{+}_{1}\rangle+\langle\hat{s}^{-}_{1}\hat{s}^{+}_{2}\hat{s}^{+}_{3}\hat{s}^{-}_{4}\rangle)+2\langle\hat{s}^{z}_{2}\hat{s}^{z}_{1}\rangle\right]\notag\\
    &=\frac{1}{4}\left(
    F \begin{bmatrix}
      +&-\\
      -&+
    \end{bmatrix}+
    2F
    \begin{bmatrix}
      +&+&-&-\\
      -&+&-&+
    \end{bmatrix}-2F
    \begin{bmatrix}
      +&+&+&-\\
      -&+&+&+
    \end{bmatrix}\right)+\frac{\Delta^2}{2}\left(
    F
    \begin{bmatrix}
      +&-\\
      -&+
    \end{bmatrix}
    -2F
    \begin{bmatrix}
      +&+&-&+\\
      +&-&+&+
    \end{bmatrix}
    \right)\notag\\
    &\quad+\frac{\Delta}{4}\left(8F
    \begin{bmatrix}
      +&+&+&-\\
      +&-&+&+
    \end{bmatrix}-4F
    \begin{bmatrix}
      +&+&-\\
      -&+&+
    \end{bmatrix}+2F
    \begin{bmatrix}
      +&-&-&+\\
      -&+&+&-
    \end{bmatrix}-2F
    \begin{bmatrix}
      +&-&+&-\\
      -&+&-&+
    \end{bmatrix}+2F
    \begin{bmatrix}
      +&+\\
      +&+
    \end{bmatrix}-F
    \begin{bmatrix}
      +\\
      +
    \end{bmatrix}\right).
\end{align}
In the last line, we rewrote correlation functions using the $F$ symbol defined in Ref.~\cite{Kato}. The analytic expressions of these correlation functions are obtained in Ref.~\cite{Kato}. Plugging in their results, we find Eq.~(9) in the main text.

\end{document}